\documentclass[9pt,twocolumn,twoside,final]{pnas-new}

\templatetype{pnasresearcharticle} 
\makeatletter
\fancypagestyle{firststyle}{
  \fancyhf{} 

}
\def\PNAS@mk@linecount{}        
\def\PNAS@linecountLO{}         
\def\PNAS@linecountRO{}         
\def\PNAS@linecountLE{}         
\def\PNAS@linecountRE{}         
\def\PNAS@mk@linecountsecpage{}
\def\PNAS@linecountsecpageLO{}
\def\PNAS@linecountsecpageRO{}
\def\PNAS@linecountsecpageLE{}
\def\PNAS@linecountsecpageRE{}
\makeatother

\usepackage{comment}
\usepackage{natbib}
\usepackage{orcidlink}

\newcommand{\apj}{The Astrophysical Journal}
\newcommand{\apjl}{The Astrophysical Journal Letters}
\newcommand{\mnras}{Monthly Notices of the Royal Astronomical Journal}
\newcommand{\nat}{Nature}
\newcommand{\aap}{Astronomy and Astrophysics}
\newcommand{\aj}{Astronomical Journal}
\newcommand{\psj}{Planetary Science Journal}
\newcommand{\icarus}{Icarus}

\begin{document}

\title{Exploring the Sub-Neptune Frontier with JWST}

\author[a,1 \orcidlink{0000-0002-4869-000X}]{Nikku Madhusudhan}
\author[a,b \orcidlink{0000-0002-0931-735X}]{M\aa ns Holmberg}
\author[a \orcidlink{0000-0001-6839-4569}]{Savvas Constantinou}
\author[a \orcidlink{0000-0001-6067-0979}]{Gregory J. Cooke}

\affil[a]{Institute of Astronomy, University of Cambridge, Cambridge, CB3 0HA, UK}
\affil[b]{Space Telescope Science Institute, Baltimore, MD 21218, USA}

\leadauthor{Madhusudhan}

\significancestatement{Sub-Neptune planets dominate the exoplanet population but have no analogues in the solar system. Sized between Earth and Neptune, the nature of such planets remains uncertain, spanning rocky gas dwarfs, mini-Neptunes and water worlds, with some {potentially} capable of harbouring habitable conditions. Sub-Neptunes are therefore a central focus in the study of planetary processes in low-mass exoplanets with many open questions. Pioneering JWST observations have led to detections of prominent molecules in several sub-Neptune atmospheres. The chemical abundances are providing initial insights into compositional trends, with important implications for their atmospheric diversity, internal structures, formation mechanisms, and habitability. These results set the stage for a generalised classification of volatile-rich sub-Neptunes and offer an early panoramic view of this exotic frontier.}

\authorcontributions{NM conceived, planned and led the research and writing of the manuscript. MH contributed to the section on JWST observations, SC contributed to the section on atmospheric retrievals, and GC contributed to {the} section on interiors and habitability. All authors contributed to editing the manuscript and producing the figures.}
\authordeclaration{The authors declare no competing interest.}
\correspondingauthor{\textsuperscript{1}To whom correspondence should be addressed. E-mail: nmadhu@ast.cam.ac.uk}

\keywords{Exoplanets $|$ Sub-Neptunes $|$ Exoplanet atmospheres $|$ Interiors $|$  Habitability $|$ JWST $|$}

\begin{abstract}

Sub-Neptune planets, with sizes and masses between those of Earth and Neptune, dominate the exoplanet population. Sub-Neptunes are expected to be the most diverse family of the exoplanet population, potentially including rocky gas dwarfs, water worlds, and mini-Neptunes, with a wide range of atmospheric, surface and interior conditions. With no analogue in the solar system, these planets open fundamental questions in planetary processes, origins, and habitability, and present new avenues in the search for life elsewhere. Atmospheric observations with the James Webb Space Telescope (JWST) are enabling unprecedented characterization of sub-Neptunes, starting with the first detections of carbon-bearing molecules in the habitable zone sub-Neptune K2-18~b. We survey the present landscape of JWST observations and atmospheric inferences of sub-Neptunes, which in turn provide key insights into their atmospheric processes, internal structures, surface conditions, formation pathways and potential habitability. The atmospheric abundance constraints reveal evidence of chemical disequilibria, and insights into the planetary mass-metallicity relation in the sub-Neptune regime. Similarly, for sub-Neptunes with H$_2$O-rich interiors, increasing atmospheric H$_2$O abundances with the equilibrium temperature may indicate the existence of a critical temperature for transition from H$_2$ dominated atmospheres with tropospheric cold traps to those with steamy atmospheres. The chemical abundances also provide initial evidence for diverse planet types, from potentially habitable hycean worlds to steam worlds with super critical water layers. These planet types serve as benchmarks for an emerging taxonomy of volatile-rich sub-Neptunes as a function of their equilibrium temperature and atmospheric extent, heralding a new era of chemical classification of low-mass exoplanets with JWST. 

\end{abstract}

\dates{This manuscript was compiled on \today}

\maketitle
\thispagestyle{firststyle}
\ifthenelse{\boolean{shortarticle}}{\ifthenelse{\boolean{singlecolumn}}{\abscontentformatted}{\abscontent}}{}

\firstpage[17]{2}

\dropcap{W}e are at the beginning of a new era in exoplanetary science. Exoplanet demographics reveal that planets between Earth and Neptune in size dominate the known exoplanet population. Sub-Neptunes typically refer to planets  with radii between $\sim$1.5-4 R$_\oplus$, smaller than Neptune but larger than rocky planets with thin atmospheres \citep{Fulton2017, rogers2015, bean_nature_2021}. Sub-Neptunes are expected to span a wide range of planet types \citep{Rogers2010a, Valencia2013, Madhusudhan2020, zeng_growth_2019}, including planets with rocky interiors and thick H$_2$-rich envelopes (gas dwarfs), smaller versions of Neptune (mini-Neptunes) with volatile-rich interiors and H$_2$-rich envelopes, and water worlds with H$_2$O-dominated interiors and varied atmospheric compositions. The water worlds can include steam worlds with H$_2$O-dominated atmospheres, ocean worlds with terrestrial-like atmospheres \citep{leger_new_2004} and hycean worlds with H$_2$-rich atmospheres \citep{madhusudhan_habitability_2021}. The central enigma underlying the sub-Neptune frontier is the lack of any such planet in the solar system which can serve {as} a reliable archetype. The abundance of sub-Neptunes has opened an unprecedented and uncharted discovery space regarding their formation mechanisms, interior and surface processes, atmospheric diversity, and potential for habitability.

A defining feature of the sub-Neptune population is the existence of a `Radius Valley' \citep{Fulton2017}, a bimodal distribution in planetary radii centered around 1.8 R$_\oplus$ with peaks near 1.4 R$_\oplus$ and 2.4 R$_\oplus$. The origin of this distribution is one of the most fundamental open questions in exoplanet science with important implications for the formation and nature of low-mass planets \citep{Mordasini2020, bean_nature_2021}. Two contesting hypotheses have been put forward to explain the distribution with different formation and evolutionary pathways. In one scenario, the two peaks in the distribution are both populated by rocky planets, but with the larger population retaining their primordial H$_2$-rich envelopes whereas the smaller ones {having lost} their envelopes through photoevaporation or core powered mass loss \citep{owen_evaporation_2017,Ginzburg2018}. In the second scenario, planets in the smaller peak are still rocky planets without significant envelopes but those in the larger peak are primarily volatile-rich planets with a large inventory (up to tens of percent) of water in the interiors along with H$_2$-rich envelopes \citep{zeng_growth_2019, Venturini2020A&A, burn2024, Luque2022}, owing to volatile accretion and migration. Each of the scenarios have implications for the internal structures and atmospheric compositions. Bulk parameters, such as mass, radius, and density, are alone not sufficient to differentiate between the different possibilities for their internal structures \citep{Rogers2010a,  Valencia2013, bean_nature_2021}. Therefore, atmospheric observations of sub-Neptunes are critical to provide deeper insights into their compositions and origins.

The sub-Neptune regime also represents a missing link in our understanding of planetary processes. In the solar system, there is a marked difference between the atmospheres of terrestrial planets (e.g. Earth, Venus and Mars) and those of ice giants (Neptune and Uranus). While the former host thin secondary atmospheres dominated by heavy molecules like N$_2$ or CO$_2$, the latter have deep H$_2$-dominated primary atmospheres with only trace amounts of {heavier} molecules like CH$_4$ and NH$_3$ \citep{Atreya2022}. As sub-Neptunes span a continuum in bulk properties between terrestrial planets and ice giants{,} their atmospheres can be expected to span a diverse range of physical and chemical processes \citep[e.g.][]{Hu_photo21, Hu2021, Yu2021, Tsai2021, Schlichting2022, kite_atmosphere_2020}, including new avenues for planetary habitability \citep{madhusudhan_habitability_2021}.  Therefore, sub-Neptune planets present a fortuitous opportunity to study a continuum of planetary processes in the low-mass regime as well as to discover key transitions and new processes in this uncharted territory. 

\begin{figure*}[t!]
\begin{center}
\includegraphics[angle=0,width=0.48\textwidth]{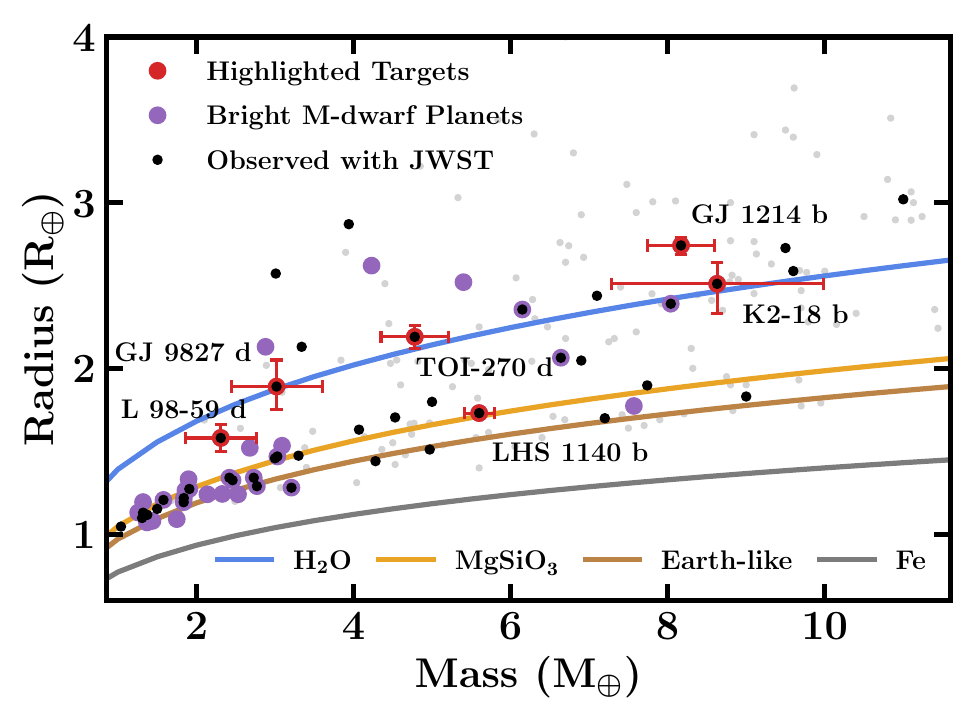}
\includegraphics[angle=0,width=0.48\textwidth]{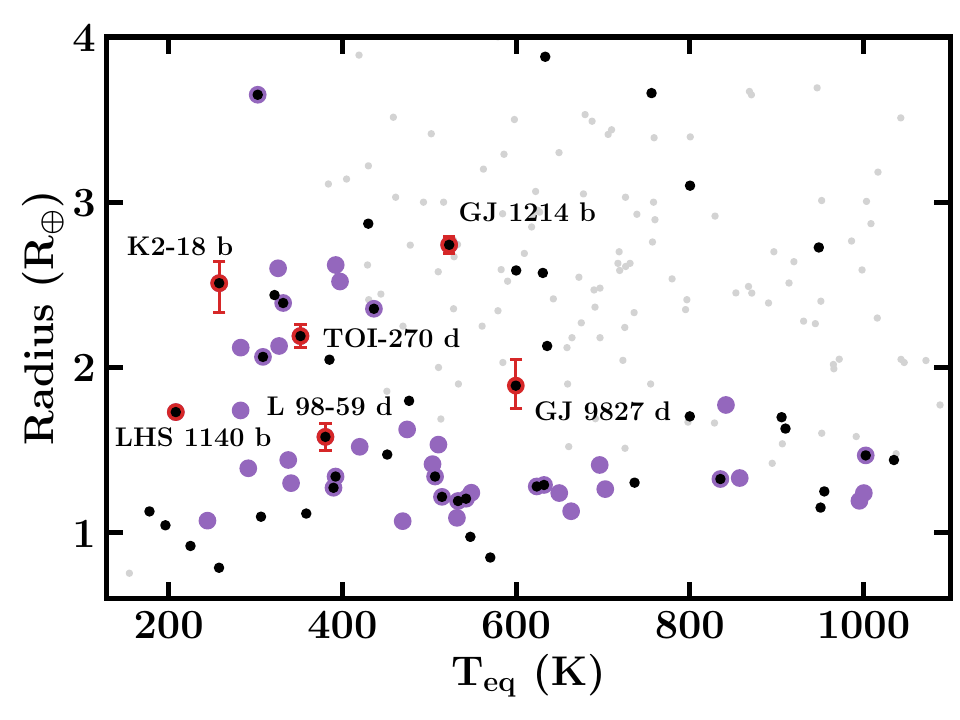}
\vspace{-3mm}
\caption{Bulk properties of promising sub-Neptunes. The black points represent targets observed with JWST, while purple points indicate sub-Neptunes orbiting bright M-dwarfs with $\mathrm{J_{mag}} < 10$ and $\mathrm{M}_* < 0.5\,\mathrm{M}_\odot$, favorable for atmospheric characterisation. Grey points denote other exoplanets with precise mass and radius measurements. {Six} targets of interest with recent JWST observations are shown in red along with the uncertainties on their masses and radii: K2-18~b \cite{Montet2015, Cloutier2019}, TOI-270~d \cite{van2021masses,MikalEvans2023}, GJ~9827~d \cite{Passegger2024}, GJ~1214~b \cite{cloutier2021}, LHS~1140~b \cite{Cadieux2024}{, and L~98-59~d} \citep{cloutier2019b}. For the equilibrium temperatures, $T_\mathrm{eq}$, we assume a Bond albedo of $A_\mathrm{B} = 0.3$ and full redistribution. {The mass--radius curves in the left panel are from \citep{Madhusudhan2020} and the planet properties are obtained from the NASA Exoplanet Archive.}}
\label{fig:population}
\end{center}
\vspace{-5mm}
\end{figure*}

Atmospheric characterisation of sub-Neptune planets has remained a formidable barrier in the pre-JWST era. Starting with the early observations with {the Hubble Space Telescope (HST)} of {a} featureless {transmission} spectrum of the sub-Neptune GJ~1214 b \cite{kreidberg2014} a decade ago, obtained with $\sim$90 hours of HST time, substantial efforts have been dedicated to such planets. Some successes with HST were seen in the {inferences} of atmospheric features in the 1.1-1.7 $\mu$m range, which overlaps with the JWST NIRISS instrument, for a few temperate sub-Neptunes orbiting bright M dwarfs such as K2-18~b \citep{Benneke2019, Tsiaras2019}, TOI-270~d \citep{MikalEvans2023}, and GJ~9827~d \citep{roy_water_2023}, also with substantial HST time in some cases, e.g. over 50 hours for K2-18~b. Whilst the observed features for K2-18~b were initially attributed to H$_2$O absorption \citep{Benneke2019, Tsiaras2019, Madhusudhan2020} with notable upper limits on CH$_4$ \citep{Madhusudhan2020}, significant degeneracies remained with other potential contributions, e.g. with CH$_4$ \citep{Blain2021,Bezard2020} and stellar contamination \citep{Barclay2021}. Furthermore, the possibility of clouds/hazes attenuating the spectral features in transmission spectroscopy was also considered to be a significant challenge for characterizing temperate atmospheres \cite[e.g.][]{kreidberg2014, Stevenson2016}. More recently, it has also been suggested that the prevalence of clouds/hazes may not have a linear dependence on temperature, with higher propensity for clouds/hazes in the $\sim$500-800 K range and clearer atmospheres for more temperate planets \cite{Yu2021_clouds, Brande2024}. HST continues to make important contributions towards understanding sub-Neptune atmospheres, especially with regards to atmospheric escape \citep[e.g.][]{Loyd2025_TOI776_Escape}. However, the limited spectral range and sensitivity available in the pre-JWST era rendered detailed characterization of sub-Neptune atmospheres challenging at a population level.

The advent of JWST has transformed this frontier overnight. With only 14 hours of observations in the first year of JWST, a near-infrared (0.8--5.2 \textmu m)  transmission spectrum of K2-18~b \citep{madhusudhan_carbon-bearing_2023} led to the first detections of carbon-bearing molecules in a temperate exoplanet, resolving the longstanding missing methane problem and the degeneracy between CH$_4$ and H$_2$O. It also provided evidence for chemical disequilibrium in a habitable zone\footnote{The habitable zone is considered to be the region around a star in which a planet can host liquid water on its surface \citep{Kasting1993, kopparapu2013, madhusudhan_habitability_2021}.} exoplanet, and important constraints on the possible internal structure, surface conditions and potential habitability. In subsequent observations, chemical constraints have also been reported for several other sub-Neptunes \citep[e.g.][]{holmberg_possible_2024, benneke_jwst_2024, piaulet2024, Damiano2024, Kempton2023, Gressier2024, Hu2024}. These observations have wide ranging implications for sub-Neptune atmospheres, interiors, formation pathways and habitability. These developments have made the study of sub-Neptunes one of the most dynamic frontiers in exoplanetary science. In this work, we present a panoramic glimpse into this exotic landscape {as observed with JWST}. 

In what follows, we present an overview of current observations of sub-Neptune planets with JWST and the resulting inferences. We subsequently discuss constraints on a range of physical and chemical processes in sub-Neptune atmospheres. We then consider constraints on their possible internal structures and surface conditions which in turn motivate a generalized classification of volatile-rich sub-Neptunes. We finally discuss the prospects for habitable conditions on temperate sub-Neptunes. 

\section*{JWST Observations}
\label{sec:observations}
JWST is revolutionising atmospheric spectroscopy of sub-Neptunes. Here we discuss atmospheric detections reported for promising targets, and early lessons on the challenges and promises with JWST.

\subsection*{Sub-Neptune Observations with JWST}

A diverse set of sub-Neptunes have been observed with JWST. Fig.~\ref{fig:population} shows a population of sub-Neptunes as a function of mass, radius and equilibrium temperature ($T_{\mathrm{eq}}$), highlighting {key} planets with JWST observations and promising targets for atmospheric characterisation. JWST detections of prominent molecules have been reported in the atmospheres of several promising sub-Neptunes, starting with the habitable zone sub-Neptune K2-18~b \citep{madhusudhan_carbon-bearing_2023}, followed by somewhat warmer planets TOI-270~d \citep{holmberg_possible_2024, benneke_jwst_2024} and GJ~9827~d \citep{piaulet2024}. {These observations utilized a combination of NIRISS SOSS and/or NIRSpec G395H. Additionally, robust detections have also been reported for a warm exo-Neptune GJ~3470~b \citep{Bonfils2012}, using NIRCam F322W2 and F444W \citep{Beatty2024}, which serves as an end-member case (being Neptune-sized), for the sub-Neptune regime. Moreover, the sub-Neptunes GJ~1214~b, LHS~1140~b, and L~98-59~d (TOI-175~d) have also been observed using transmission spectroscopy with JWST \citep{schlawin_possible_2024, Damiano2024, Cadieux2024b, Gressier2024}, with some evidence for atmospheric features. Figs.~\ref{fig:spectra_niriss} and \ref{fig:spectra_nirspec} illustrate the recent transmission spectra obtained for these targets.

\begin{figure}[t!]
\begin{center}
\includegraphics[angle=0,width=0.48\textwidth]{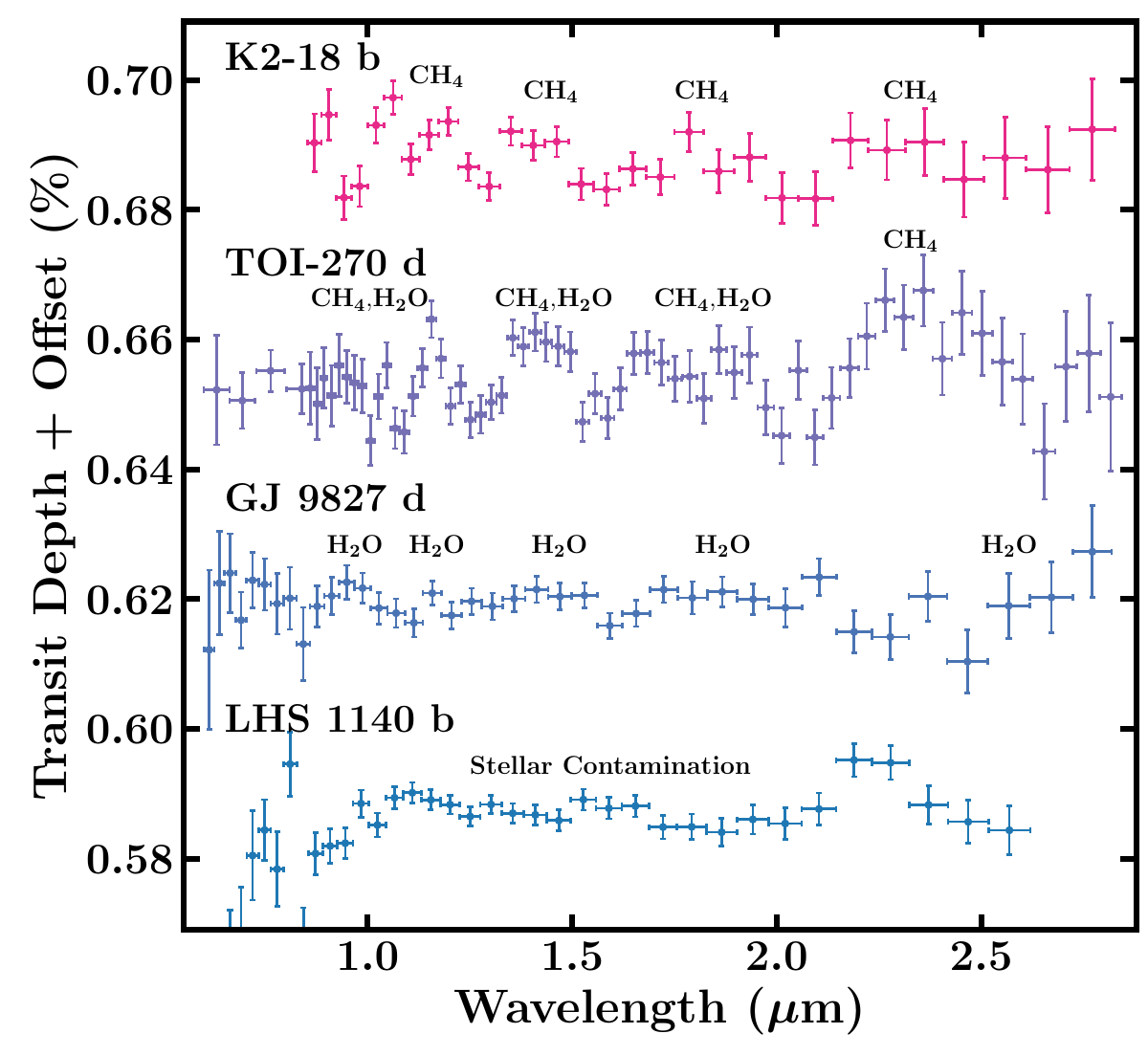}
\vspace{-5mm}
\caption{{JWST transmission spectra of four sub-Neptunes in the 0.6-2.8~$\mu$m wavelength range obtained with NIRISS SOSS. These spectra of K2-18~b, TOI-270~d, GJ~9827~d and LHS 1140~b are obtained from ref.~\cite{madhusudhan_carbon-bearing_2023},  ref.~\citep{benneke_jwst_2024}, ref.~\cite{piaulet2024}, and ref.~\cite{Cadieux2024b}, respectively. For the transmission spectra of GJ~9827~d and LHS 1140~b, we binned every four spectral points for visual clarity. The spectral features of prominent molecules as inferred in the respective works are labeled.}
}
\label{fig:spectra_niriss}
\end{center}
\vspace{-4mm}
\end{figure}

\begin{figure}[t!]
\begin{center}
\includegraphics[angle=0,width=0.48\textwidth]{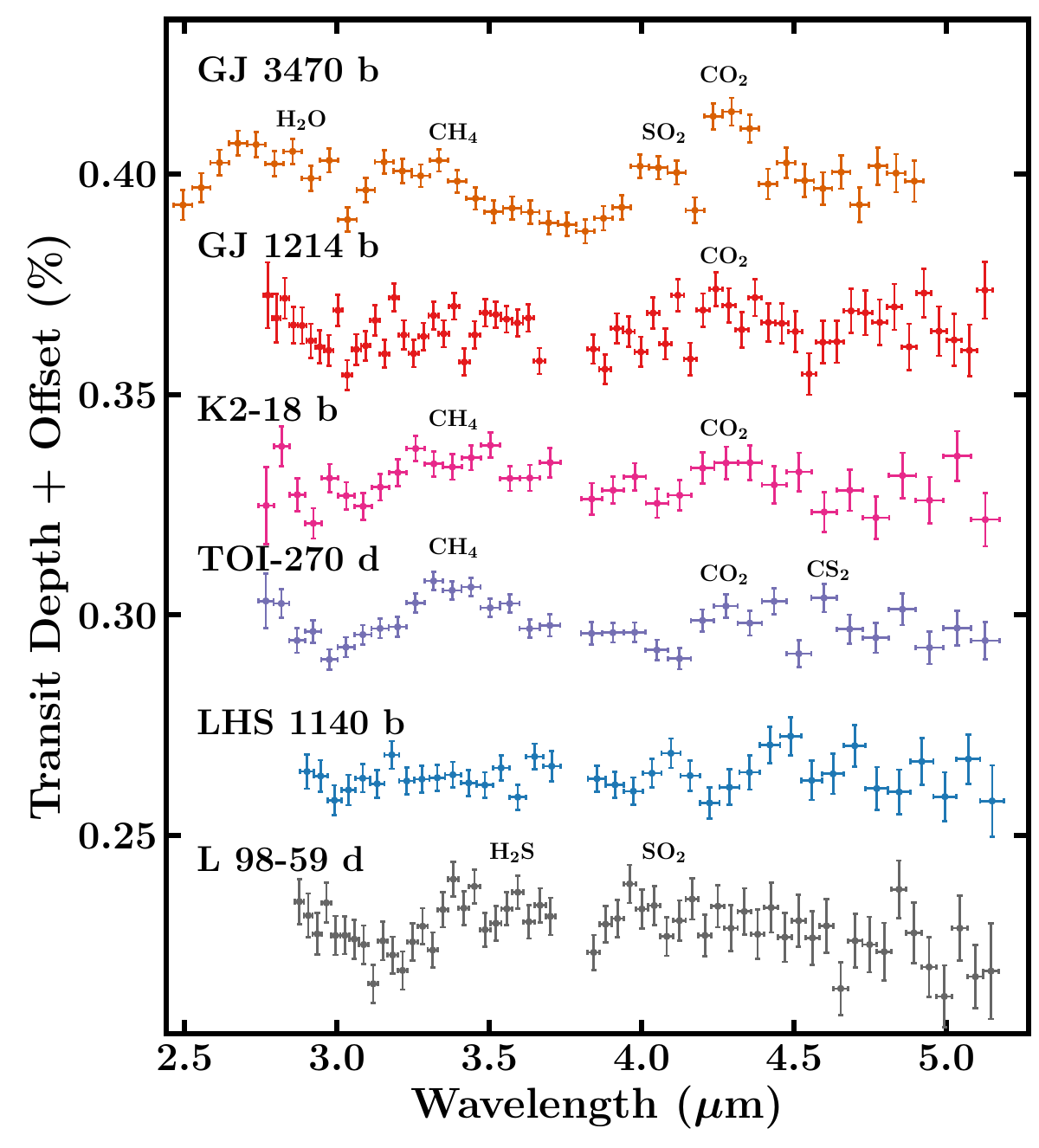}
\vspace{-5mm}
\caption{{JWST transmission spectra of six sub-Neptunes in the 2.5-5.2~$\mu$m wavelength range obtained with NIRSpec G395H or NIRCam F322W2/F444W. The spectra of GJ~3470~b, GJ~1214~b, K2-18~b, TOI-270~d, LHS 1140~b, and L~98-59~d are obtained from ref. \cite{Beatty2024}, ref. \cite{schlawin_possible_2024}, ref. \cite{madhusudhan_carbon-bearing_2023}, ref. \cite{holmberg_possible_2024}, ref. \cite{Damiano2024}, and ref. \cite{Gressier2024}, respectively. In the case of GJ~3470~b, we binned every four spectral points for visual clarity. The spectral features of prominent molecules as inferred in the respective works are labeled. The LHS~1140~b spectrum has been inferred as showing no prominent spectral features \citep{Damiano2024}, instead favoring a high-mean-molecular-weight atmosphere.} 
}
\label{fig:spectra_nirspec}
\end{center}
\vspace{-8mm}
\end{figure}

The coolest of these targets, and the first to be observed with JWST, is K2-18~b, a habitable zone sub-Neptune with an equilibrium temperature of $T_{\mathrm{eq}} = 258$ K (assuming $A_\mathrm{B} = 0.3$ and full redistribution). The JWST transmission spectrum of K2-18~b revealed prominent spectral features of CH$_4$ and CO$_2$ \cite{madhusudhan_carbon-bearing_2023} -- marking the first detection of carbon-bearing molecules in a habitable zone exoplanet.  With a higher temperature of $T_{\mathrm{eq}} = 352$ K, TOI-270~d is another temperate sub-Neptune observed with JWST. The transmission spectrum of TOI-270~d led to robust detections of CH$_4$ and CO$_2$, with further evidence for CS$_2$ and a potential inference of H$_2$O \cite{holmberg_possible_2024, benneke_jwst_2024}. Similar to K2-18~b, NH$_3$ and CO were not detected in the atmosphere of TOI-270~d. However, the spectrum of TOI-270~d did not provide strong evidence for clouds/hazes, in contrast to K2-18~b, which showed a $\sim$$3 \sigma$ preference for the {presence} of clouds/hazes. Additionally, the transmission spectrum of  GJ~9827~d, a hotter sub-Neptune with a temperature of $T_{\mathrm{eq}} = 600$ K, showed spectral features of mainly H$_2$O by combining two visits with NIRISS SOSS \cite{piaulet2024}, although with less prominent spectral signatures compared to K2-18~b and TOI-270~d. Finally, the JWST spectrum of {exo-Neptune} GJ~3470~b, with an equilibrium temperature of $T_{\mathrm{eq}} = 634$ K, exhibits prominent features of H$_2$O, CH$_4$, CO$_2$, and SO$_2$ \cite{Beatty2024}.

Besides these four targets, several other sub-Neptune exoplanets have been observed with JWST, {as mentioned above,} such as LHS~1140~b \cite[{NIRISS SOSS, NIRSpec G235H/G395H;}][]{Damiano2024, Cadieux2024b}, GJ~1214~b \cite[{MIRI LRS, NIRSpec G395H;}][]{Kempton2023, schlawin_possible_2024}, L~98-59~d \cite[{NIRSpec G395H;}][]{Gressier2024, Banerjee2024} and 55~Cancri~e \cite[{in emission, using MIRI LRS and NIRCam F444W;}][]{Hu2024, Patel2024}. {Ref~\cite{Damiano2024} and ref~\cite{Cadieux2024b} found evidence against a H$_2$-rich atmosphere on the habitable zone sub-Neptune LHS~1140~b, with the data favouring a high mean molecular weight (MMW) atmosphere with additional contribution from unocculted stellar heterogeneities at shorter wavelengths. For GJ~1214~b, a warm sub-Neptune found to host an atmosphere dominated by clouds, \citep{kreidberg2014, Kempton2023}, CO$_2$ and CH$_4$ were tentatively inferred \citep{schlawin_possible_2024, Ohno2025}. However, the robustness of these detections is hindered by the possibility of clouds/hazes. Similarly, atmospheric spectroscopy of L~98-59~d, a sub-Neptune slightly cooler than GJ~1214~b has revealed tentative indications of sulfur-bearing species H$_2$S and SO$_2$ \citep{Gressier2024, Banerjee2024}, with these molecules potentially resulting from tidal-driven volcanism \citep{Seligman2024Tidal}. Lastly, 55~Cancri~e, a strongly irradiated sub-Neptune, with $T_{\mathrm{eq}} \sim 2000$ K, was found to potentially host an atmosphere, with indications of it primarily being comprised of CO and CO$_2$ \citep{Hu2024}. Further observations of these targets can confirm the above findings which for now remain tentative. Moreover, the above list will undoubtedly expand as more sub-Neptunes across the temperature range are presently under observation with JWST.} Other JWST studies of sub-Neptunes reveal seemingly featureless spectra \cite[e.g.,][]{May2023, Wallack2024}. Overall, these results highlight the diversity of sub-Neptune exoplanets, emerging trends and early lessons. 

\subsection*{Early Lessons}

\begin{figure}[t!]
\begin{center}
\includegraphics[angle=0,width=0.48\textwidth]{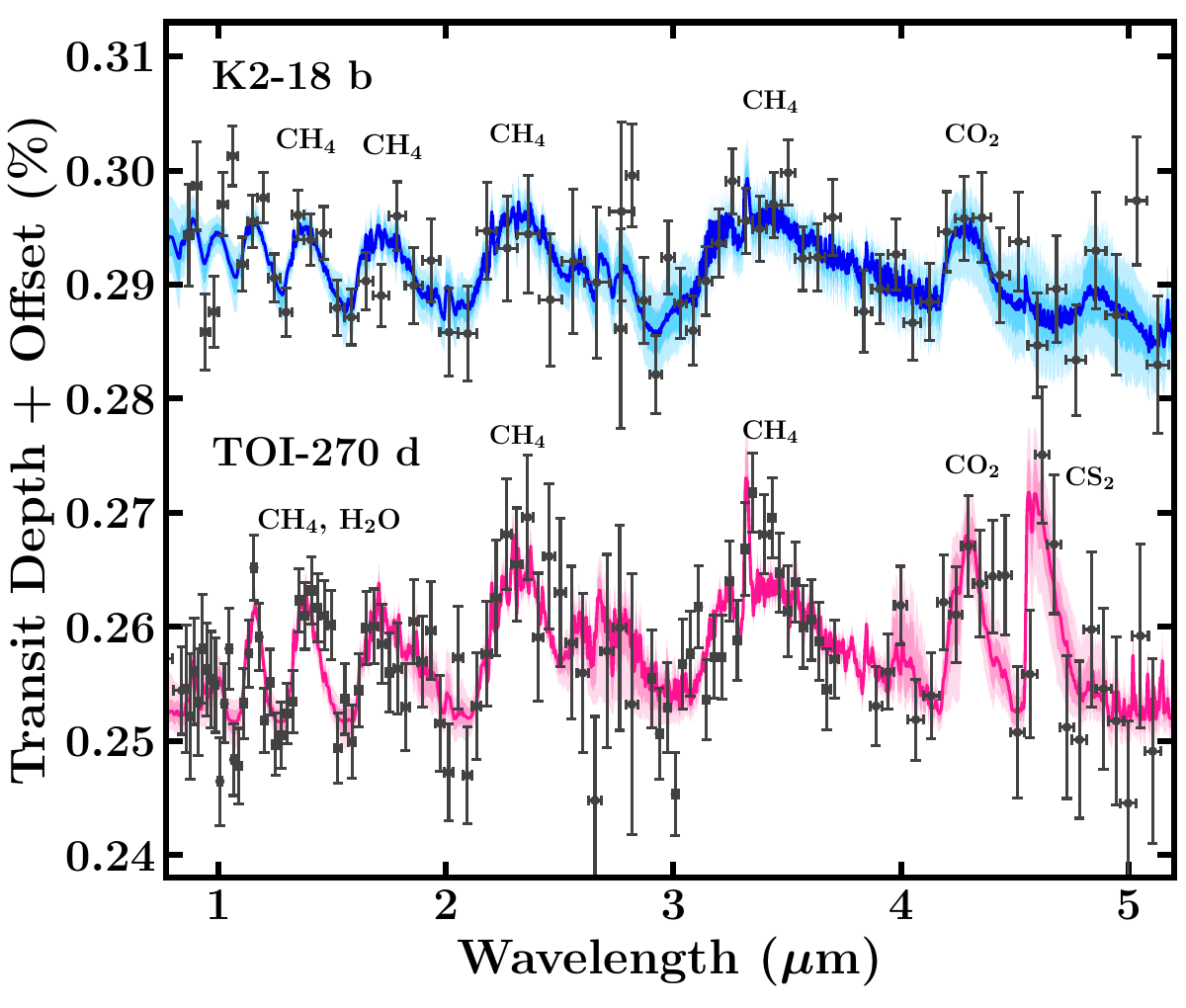}
\vspace{-5mm}
\caption{JWST transmission spectra of K2-18~b {and TOI-270~d,} observed with NIRISS SOSS and NIRSpec G395H. The K2-18~b and TOI-270~d data are obtained from ref. \cite{madhusudhan_carbon-bearing_2023} and ref.~\cite{benneke_jwst_2024}, respectively, and shown in grey squares with error bars. The colored contours show nominal model fits to the data, with the dark curves denoting the median retrieved model spectra, while the lighter contours denote the 1$\sigma$ and 2$\sigma$ intervals. Molecules with prominent spectral features are labeled. }
\label{fig:spectra}
\end{center}
\vspace{-4mm}
\end{figure}

 JWST Cycles 1-3 have dedicated around 800 hours to transit spectroscopic observations of sub-Neptunes across more than 40 unique targets. However, only a subset of observed exoplanets has resulted in robust atmospheric detections, as described above. Specifically, published JWST observations of K2-18~b, TOI-270~d, GJ~9827~d, and GJ~3470~b amount to only about 110 hours so far. These observations provide important insights for successful atmospheric characterisation of sub-Neptunes using JWST.

One of the early identified challenges for atmospheric characterization of small exoplanets is the presence of stellar heterogeneities and flares. These factors can contaminate the observed transmission spectrum, complicating the analysis. This issue has proven particularly significant for JWST observations of exoplanets orbiting late M-dwarfs, as reported for TRAPPIST-1 \cite{Lim2023, radica2025promise} and other systems \cite[e.g.,][]{moran_high_2023}. Successful JWST observations of mid-M dwarfs, such as K2-18 and TOI-270 {(see Fig.~\ref{fig:spectra})}, clearly demonstrate that low stellar activity is essential for mitigating these effects. 

Additionally, the temperature of the planet plays a crucial role in its observability, all other factors remaining the same. For planets with a large H$_2$O content, high temperatures can lead to a mixed {H$_2$-H$_2$O} envelope with high MMW \cite{Nixon21_ocean, benneke_jwst_2024}, as described in the \textit{Sub-Neptune Interiors and Surface Conditions} section, reducing the overall amplitude of the spectral feature. {At the same time, both observations and theoretical studies also seem to indicate that planets with $T_\mathrm{eq}$ below $\sim$500 K are less likely to be affected by clouds/hazes attenuating transmission spectra \citep{madhusudhan_carbon-bearing_2023, holmberg_possible_2024, Yu2021_clouds, Brande2024}. Based on the above findings, temperate} sub-Neptunes with $T_\mathrm{eq} \lesssim 500$~K {orbiting relatively quiet stars} appear ideal {for atmospheric characterisation}. For optimal results, it is also vital that the host star is bright enough, e.g. $J < 10$ \citep{madhusudhan_habitability_2021}. These considerations make target selection particularly important for reliable atmospheric characterisation of sub-Neptunes.

Moreover, several systematics have been identified to be important for transit observations of small planets with JWST. Notable among these are offsets or broadband variations possible in the transmission spectrum across different instruments, detectors, and visits \cite{moran_high_2023, madhusudhan_carbon-bearing_2023, piaulet2024}. Key factors that may contribute to such variations include treatment of limb darkening, light-curve detrending, stellar variability and possible instrumental effects \citep{May2023, holmberg_possible_2024, madhusudhan_carbon-bearing_2023, Cadieux2024b}. Some of these effects, such as offsets or stellar heterogeneities, can be considered in the atmospheric retrievals \citep{May2023, madhusudhan_carbon-bearing_2023} while also being potentially degenerate with some of the atmospheric model parameters. Additionally, correlated noise is another potential concern affecting some sub-Neptune observations \cite[e.g.,][]{Kirk2024, Patel2024, Wallack2024}. This requires careful attention to the treatment of noise properties{, with multiple transits preferred for robustness \citep{May2023}}. Overall, it is crucial to thoroughly investigate the effects of various data analysis assumptions and approaches to ensure robust inference and interpretation.

\section*{Atmospheric Retrievals} 
\label{sec:retrieval}

The JWST spectra discussed above are providing the first detailed constraints on the atmospheric properties of several sub-Neptunes using atmospheric retrieval methods \citep{madhusudhan2018atmospheric}. As most of the chemical detections for sub-Neptunes to date have been made using transmission spectroscopy, the retrieved properties pertain to the observable atmosphere at the day-night terminator region. The retrieved properties include volume mixing ratios of prominent molecular species along with constraints on the temperature structure, the presence of clouds/hazes, and the effect of stellar contamination.

Here we focus on the atmospheric constraints for three sub-Neptune targets for which high-confidence molecular detections have been reported {with} high-precision JWST { transmission spectra.} These include K2-18~b \citep{madhusudhan_carbon-bearing_2023}, TOI-270~d \citep{holmberg_possible_2024, benneke_jwst_2024} and GJ~9827~d \citep{piaulet2024}. We also discuss the retrieved constraints for the exo-Neptune GJ~3470~b \citep{Beatty2024} for a comparative assessment between the sub-Neptune targets and a hot ice-giant analogue in size. The retrieved chemical abundances of prominent molecules from published works for these targets are shown in Fig.~\ref{fig:abundance_trends}. 

\subsection*{Atmospheric Constraints with JWST Observations}

The high-precision JWST spectra have led to precise abundance constraints for several prominent molecules. For K2-18~b, the observations led to robust detections of prominent carbon-bearing molecules methane (CH$_4$) and carbon dioxide (CO$_2$) in a H$_2$-rich atmosphere with retrieved abundances of log(CH$_4$) = $-1.74^{+0.59}_{-0.69}$ and log(CO$_2$) = $-2.09^{+0.51}_{-0.94}$ \citep[one-offset case;][]{madhusudhan_carbon-bearing_2023}. The observations did not show evidence for other prominent molecules such as H$_2$O, NH$_3$ and CO, and provided 2$\sigma$ upper limits of $-3.06$, $-4.51$ and $-3.50$, respectively, on their log mixing ratios. The retrievals obtained nominal constraints on the presence of clouds/hazes, which were preferred at a 3$\sigma$ level. The photospheric (10 mbar) temperature at the terminator was constrained to be 242$^{+79}_{-57}$ K, consistent with the non-detection of H$_2$O which would be condensed out. 

JWST observations of the temperate sub-Neptune, TOI-270~d, led to similar atmospheric detections and abundance constraints of carbon-bearing molecules as for K2-18~b, from two recent independent studies \citep{holmberg_possible_2024, benneke_jwst_2024}. As discussed above, both works reported confident detections of CH$_4$ and CO$_2$, and potential inferences of CS$_2$ and H$_2$O. The abundance constraints reported by these works range between: log(CH$_4$) = $-2.72^{+0.41}_{-0.50}$, log(CO$_2$) = $-2.46^{+0.71}_{-0.92}$, log(H$_2$O) =$ -1.91^{+0.57}_{-0.94}$, log(CS$_2$) = $-3.07^{+0.74}_{-0.91}$ \citep[dual transit case;][]{holmberg_possible_2024}, and log(CH$_4$) = $-1.64^{+0.38}_{-0.36}$, log(CO$_2$) = $-1.67^{+0.40}_{-0.60}$, log(H$_2$O) =$ -1.10^{+0.31}_{-0.92}$, log(CS$_2$) = $-3.44^{+0.66}_{-0.67}$ \citep{benneke_jwst_2024}. Similarly to K2-18~b, no evidence was found for NH$_3$ or CO in TOI-270~d.

JWST observations were also reported for a third, significantly warmer, sub-Neptune, GJ~9827~d \citep{piaulet2024}. The spectrum showed strong evidence for a H$_2$O-dominated atmosphere, with a 2$\sigma$ lower limit of log(H$_2$O) $>$ -0.5, i.e., volume mixing ratio above 31.6\%. The very high H$_2$O abundance, unlike the temperate sub-Neptunes discussed above, leads to significantly smaller spectral amplitudes despite the higher temperature ($T_{\rm eq}$ = 600 K).

Finally, the warm exo-Neptune GJ~3470~b was recently observed with JWST NIRCam \citep{Beatty2024}, leading to above 3$\sigma$ detections of H$_2$O, CH$_4$, CO$_2$, and SO$_2$ in a H$_2$-rich atmosphere \citep{Beatty2024}. The abundances were reported to be log(H$_2$O) = $-1.08^{+0.43}_{-0.52}$, log(CH$_4$) = $-4.05^{+0.25}_{-0.27}$, log(CO$_2$) = $-2.47^{+0.61}_{-0.43}$, and log(SO$_2$) = $-3.57^{+0.26}_{-0.25}$. Additionally, ref. \cite{Beatty2024} also reported the abundance constraint of CO, albeit with a lower detection significance of 1.5$\sigma$, to be log(CO) = $-0.96^{+0.18}_{-0.70}$. This means that CO is potentially the most dominant carbon-bearing species in GJ~3470~b's atmosphere, {consistent} with expectations from {vertical mixing in the atmosphere.}

\begin{figure*}[t!]
\begin{center}
\includegraphics[angle=0,width=\textwidth]{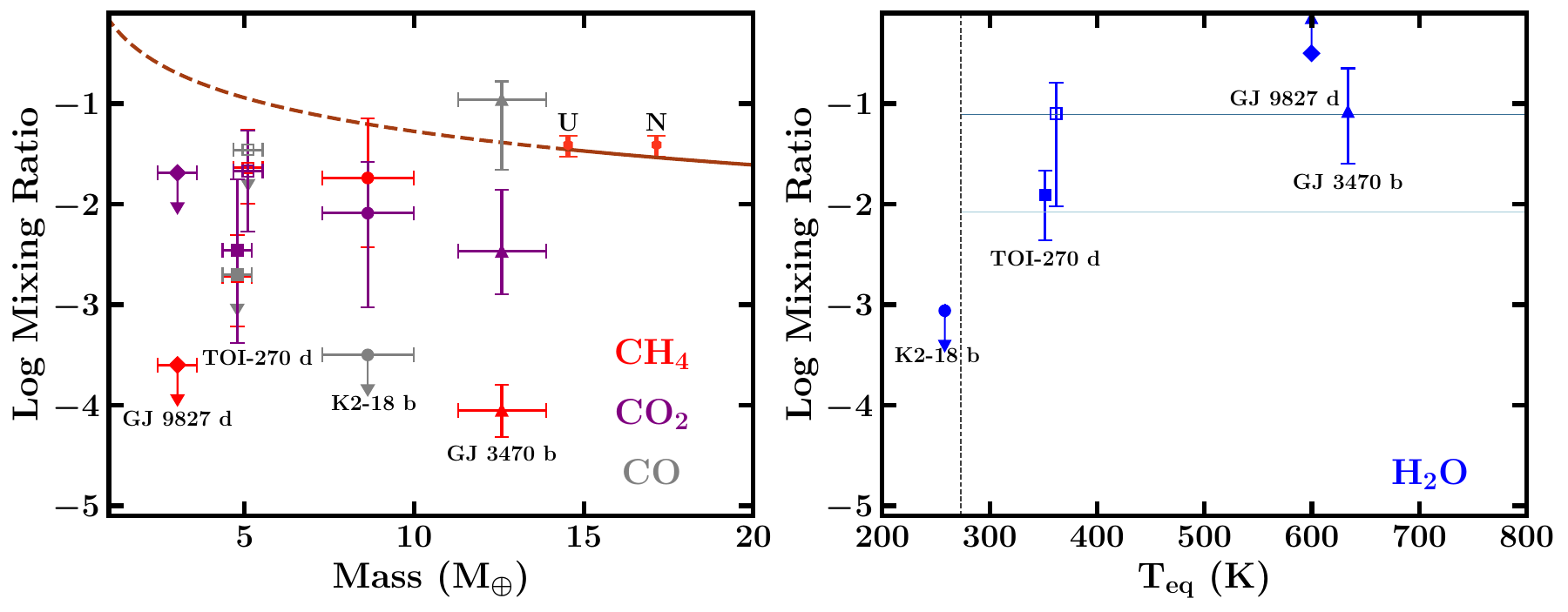}
\vspace{-5mm}
\caption{Composition trends in sub-Neptune atmospheres. Points with errorbars denote the median and 1$\sigma$ uncertainties in retrieved abundances of prominent molecules, while arrows denote a 2$\sigma$ upper {or lower} limit. The constraints are shown for four planets: K2-18~b \citep{madhusudhan_carbon-bearing_2023}, TOI-270~d \citep{holmberg_possible_2024, benneke_jwst_2024}, GJ~9827~d \citep{piaulet2024} and GJ~3470~b \citep{Beatty2024}. For TOI-270~d we show abundances reported in two independent works, denoted by filled squares \citep{holmberg_possible_2024} and open squares \citep{benneke_jwst_2024}. {\bf Left:} CH$_4$ (red) CO$_2$ (purple) and CO (grey) constraints against planetary mass, along with CH$_4$ constraints for Uranus and Neptune \cite{Atreya2022}. Also shown is a fitted mass-metallicity trend for solar system giant planets (including Saturn and Jupiter) as a solid brown line, and its extrapolation to the sub-Neptune regime as a dashed brown line. {\bf Right:} H$_2$O {mixing ratio} constraints for the same four targets against their equilibrium temperatures assuming a 0.3 Bond albedo. {Here, the mixing ratio of a molecule is the ratio of its number density relative to the total number density.} Light and dark blue lines denote the gaseous H$_2$O volume mixing {ratio} for atmospheres with a 10x and 100x solar oxygen abundance, respectively, assuming all oxygen is in H$_2$O. The vertical black dashed line indicates the 273~K condensation temperature for H$_2$O under tropospheric conditions for illustrative purposes.}
\label{fig:abundance_trends}
\end{center}
\vspace{-5mm}
\end{figure*}

As discussed in section {\it JWST Observations}, nominal chemical signatures have also been inferred in some other sub-Neptunes with preliminary constraints on their abundances. For example, a tentative inference of CH$_4$ and CO$_2$ was reported for the warm sub-Neptune GJ 1214~b, with log(CH$_4$) = $-2.25^{+0.72}_{-0.45}$ in a CO$_2$-dominated atmosphere of $91^{+8}_{-45}$\% by volume \cite{schlawin_possible_2024, Ohno2025}. Given the low combined significance of CH$_4$ and CO$_2$, of 2-3 $\sigma$, these studies suggest that more observations are required to detect both molecules independently and derive robust abundance estimates. Similarly, a sulphur-rich atmosphere was inferred for the sub-Neptune L~98-59~d, with a combination of H$_2$S and SO$_2$ reported at 2-4~$\sigma$ significance across multiple studies \cite{Banerjee2024, Gressier2024}. The abundances are constrained to log(H$_2$S) = $-0.74^{+0.14}_{-0.49}$ by ref. \cite{Gressier2024} and log(H$_2$S) = $-0.62^{+0.61}_{-6.73}$ by ref. \cite{Banerjee2024}. Additionally, ref. \cite{Banerjee2024} infer log(SO$_2$) = $-2.35^{+2.05}_{-5.88}$, while ref. \cite{Gressier2024} obtain a bimodal inference constraining log(SO$_2$) to either above -1.0 or below -1.5.

Overall, atmospheric abundance constraints have been obtained for a small sample of promising sub-Neptunes to date with more observations for a larger target sample underway. These atmospheric constraints combined with self-consistent theoretical models can provide important insights into various atmospheric processes, interior and surface conditions, formation mechanisms and potential for habitability as detailed below.

\section*{Atmospheric Processes}

The most direct inferences derived from retrieved atmospheric constraints are those of atmospheric processes{.} Such processes result in characteristic abundances of specific chemical species or groups of species relative to expectations from thermochemical equilibrium. Therefore, the atmospheric constraints can provide insights into chemical disequilibrium processes, temperature structures, clouds/hazes, atmospheric metallicity, and atmospheric extent. The atmospheric constraints at the population level can also enable key insights into planetary formation conditions and mechanisms, based on trends in atmospheric compositions {with system} properties including planet mass and irradiation.

\subsection*{Chemical Disequilibrium and Atmospheric Extent}

All planetary atmospheres in the solar system are in chemical disequilibrium { \cite{yung1999photochemistry}.} Low temperature atmospheres are particularly susceptible to chemical, radiative, and dynamical processes which can drive the atmosphere out of thermochemical equilibrium \citep{Moses2013,Madhu2016}. For example, deep H$_2$-rich atmospheres in chemical equilibrium are expected to be abundant in CH$_4$ and NH$_3$ as the dominant carriers of C and N, respectively, at low temperatures, and are replaced with CO and N$_2$ at higher temperatures. At 1 bar, the CO-CH$_4$ transition happens at $\sim$1200 K and that for N$_2$-NH$_3$ at $\sim$700 K \citep{burrows1999, Lodders2002, Moses2013}. However, vertical mixing can dredge up CO from the hotter lower atmosphere to the observable upper atmosphere even in temperate planets \citep{Yu2021, wogan_jwst_2024}. On the other hand, photochemical processes can result in destruction of CH$_4$ and NH$_3$ and enhancement of CO$_2$ and HCN. In addition, the presence of a shallow surface underlying a thin H$_2$-rich atmosphere can significantly affect the atmospheric composition by curtailing the thermochemical recycling from the lower atmosphere \citep{Yu2021, Hu2021, Tsai2021, Madhusudhan_chem_2023}. H$_2$O is expected to be the dominant oxygen-bearing molecule in temperate H$_2$-rich atmospheres but can be limited by condensation in a cool atmosphere or enhanced in the presence of a warm ocean surface or a metal-rich interior \cite[e.g.][]{Madhusudhan_chem_2023, benneke_jwst_2024}. 

The retrieved atmospheric abundances of sub-Neptunes with JWST reveal diverse disequilibrium processes in their atmospheres and potential surface-atmosphere interactions. The first JWST detection of chemical disequilibrium in a temperate sub-Neptune was made for K2-18~b \citep{madhusudhan_carbon-bearing_2023}. The high retrieved abundances of CH$_4$ and CO$_2$, and non-detection of NH$_3$ and CO simultaneously are incompatible with expectations from thermochemical equilibrium at the retrieved photospheric temperatures below 300 K. While the CH$_4$ abundance alone is consistent with up to $\sim$150$\times$solar metallicity in chemical equilibrium, the remaining abundances still need some form of chemical disequilibrium. However, chemical disequilibrium alone cannot explain the abundances if a deep H$_2$-rich atmosphere is assumed \citep{Cooke2024,Rigby_magma_2024}, necessitating the consideration of a thin H$_2$-rich atmosphere over an ocean surface, as discussed further in the \textit{Sub-Neptune Interiors and Surface Conditions} section. The similar abundance pattern observed for TOI-270~d \citep{benneke_jwst_2024, holmberg_possible_2024} presents a second instance of chemical disequilibrium in a temperate sub-Neptune with varied implications for the interior depending on the retrieved metallicity, as discussed below. 

The chemical abundances of the exo-Neptune GJ 3470 serve as an important reference for interpreting sub-Neptune compositions \citep{Beatty2024}. A limiting degeneracy in the interpretation of a sub-Neptune composition is whether the planet has a thin or deep H$_2$-rich atmosphere, e.g. as in the case of K2-18~b discussed above. In the case of GJ~3470~b, the mass and radius of the planet are comparable to Uranus and Neptune, and, hence, require a deep H$_2$-rich envelope in the interior. This breaks the degeneracy with regard to the atmospheric extent. The retrieved atmospheric abundance constraints of high H$_2$O and CO$_2$, low CH$_4$, and high CO, albeit with marginal detection significance, are consistent with expectations of a warm and deep H$_2$-rich atmosphere with vertical mixing \citep{Beatty2024}. The non-detection of NH$_3$ in this case is also consistent with the warm atmosphere.

\subsection*{Clouds/hazes and MMW} \label{sec:clouds}

In the pre-JWST era, it was expected that clouds/hazes would mute spectral features of temperate exoplanets, as exemplified by the featureless HST transmission spectrum of GJ~1214~b {in the 1-1.7 $\mu$m range} \cite{kreidberg2014}. Other studies have also discussed trends related to clouds/hazes in sub-Neptunes {suggesting an intermediate temperature range of $\sim$500-700~K, where spectral attenuation due to clouds/hazes may be expected to be most significant \citep{Yu2021_clouds,Brande2024}}. {As discussed above,} JWST observations have successfully detected atmospheric features in the spectra of several low-mass exoplanets across a wide range of conditions, with temperatures ranging from $T_\mathrm{eq} \lesssim 300$~K for K2-18~b to $T_\mathrm{eq} \sim 600$~K for GJ~9827~d and GJ~3470~b. {There have also been inferences of clouds/hazes in some sub-Neptunes, such as GJ~1214~b with $T_\mathrm{eq} \sim 550-600$~K \citep{Kempton2023, Ohno2025, schlawin_possible_2024} and K2-18~b \citep{madhusudhan_carbon-bearing_2023}, while still allowing for identification of significant spectral features. On the other hand}, a strong scattering slope suggested for GJ~3470~b using HST \citep{Benneke2019a} was found to be inconsistent with recent JWST spectra, which showed strong features of prominent molecules \citep{Beatty2024}. { It is generally the case, however, that more temperate sub-Neptunes, e.g. $T_\text{eq}$ below 400 K, are showing stronger spectral features compared to their warmer counterparts for comparable masses and radii as demonstrated for K2-18~b and TOI-270~d discussed above. More precise JWST observations of sub-Neptunes over a broader planetary temperature range would be instrumental for robustly constraining any underlying trends regarding the prevalence and nature of clouds/hazes in the sub-Neptune regime. }

An emerging alternative explanation for muted spectral features in {warmer} sub-Neptunes is that {their} high temperatures increase the MMW which, in turn, decreases the spectral amplitudes. For example, as shown in Fig.~\ref{Sub-Neptune schematic figure}, for {volatile-rich} sub-Neptunes with $T_\mathrm{eq} \gtrsim 350$~K, the atmospheres are expected to be {increasingly} dominated by H$_2$O, either as largely steam or a mixture of abundant H$_2$O and H$_2$. This would result in an increased MMW, leading to diminished spectral features. The ratio of H$_2$O to H$_2$ would depend on the amount of H$_2$ present from formation and evolution. Consequently, smaller planets with lower H$_2$ abundances are expected to have predominantly steam atmospheres, as observed for GJ~9827~d, {whereas planets with inherently lower metallicities and low O/H ratios could have lower H$_2$O abundances}. Ultimately, the detection of atmospheric features will depend on both the presence of clouds/hazes and the MMW. Therefore, low-temperature sub-Neptunes with equilibrium temperatures { below $\sim$400-500 K} \citep{Brande2024}, are likely to be optimal targets for atmospheric characterization for given bulk properties.

\subsection*{Mass-Metallicity Relation in the Sub-Neptune Regime}
The retrieved abundance constraints are beginning to allow comparative planetology in the sub-Neptune regime. The carbon abundance has served as a unique probe of planet formation in the solar system. The dominant form of carbon in the atmosphere of all solar system giant planets is CH$_4$, which shows increasing abundance with decreasing planet mass \citep{Atreya2022}. This trend has been used as an indicator of giant planet formation by core accretion \citep{Atreya2022}. The CH$_4$ abundances in the atmospheres of the ice giants Uranus and Neptune, with masses of 14.5 M$_\oplus$ and 17.2 M$_\oplus$, respectively, are at a few percent volume mixing ratio.  An extrapolation of the solar system CH$_4$ trend to lower masses predicts volume mixing ratios that are systematically higher for sub-Neptunes.

The abundances of cabon-bearing species derived for sub-Neptunes to date do not show strong evidence for increasing C/H with decreasing mass in this regime, i.e. below 10 M$_\oplus$, as shown in Fig. \ref{fig:abundance_trends}. Currently, abundance estimates of carbon-bearing molecules have been published for only two sub-Neptunes, K2-18~b \citep{madhusudhan_carbon-bearing_2023} and TOI-270~d \citep{holmberg_possible_2024}, where CH$_4$ and CO$_2$ have been detected. Accounting for the total carbon abundance by combining CH$_4$ and CO$_2$ still somewhat  underpredicts the C/H ratio of TOI-270~d relative to the solar system trend, while K2-18~b is consistent. On the other hand, the retrieved carbon abundance for the exo-Neptune GJ~3470~b \citep{Beatty2024}, which is close to Uranus and Neptune in mass, is consistent with the solar system trend; we note however that the detection significance of CO in this planet is relatively weak \citep{Beatty2024}. Finally, no carbon-bearing molecule has been detected in GJ~9827~d, the smallest exoplanet in this sample \citep{piaulet2024}. These initial estimates may indicate a carbon cliff or plateau in the mass-metallicity relation in the sub-Neptune regime. However, given the limited data available for only a few systems currently, it is too early to make a robust assessment of the trend. Nevertheless the ability to make such abundance measurements with JWST means that these initial indications can be tested with more precise abundance measurements for more planets in the future. Such measurements would help further refine the mass-metallicity relation and its implications for planetary formation in the sub-Neptune regime. On the other hand, Uranus, for which nitrogen abundance measurements are available, has nearly the solar abundance value of NH$_3$ \citep{Atreya2022}, which is not detected in K2-18~b and TOI-270~d despite the favourable low temperatures where it would be expected for a deep atmosphere. These abundances potentially indicate a different internal structure and formation pathway for these temperate sub-Neptunes compared to the ice giants in the solar system. 

\subsection*{Cold Trap and Critical Temperature}

{The atmospheric H$_2$O abundance places important constraints on the atmospheric temperature structure and interior conditions. As shown in Fig.~\ref{fig:abundance_trends}, the retrieved H$_2$O abundance is found to increase with the equilibrium temperature of the planet, from a non-detection in the case of K2-18~b with $T_{\rm eq}$ below 300 K to over $\sim$30\% for GJ~9827~d with a $T_{\rm eq}$ of 600 K. The observed trend is consistent with the presence of a cold trap in the atmosphere of K2-18~b whereby water is condensed at the tropopause leading to a dry stratosphere. On the other hand, for hotter sub-Neptunes the lack of a cold trap leads to observable water vapour in the atmosphere. Furthermore, a hot water rich interior will have H$_2$O in supercritical phase in which H$_2$ would be highly soluble, leading to a mixed envelope of H$_2$ and H$_2$O \citep{Nixon21_ocean, gupta2025miscibility}. We refer to such planets as supercritical mini-Neptunes, as discussed below in section on \textit{Classification of Volatile-rich Sub-Neptune Interiors}; these have also been referred to as mixed envelope \citep{Nixon21_ocean} or miscible envelope \citep{benneke_jwst_2024} sub-Neptunes. The high H$_2$O mixing ratio observed for GJ~9827~d is consistent with this picture and can therefore be a supercritical mini-Neptune or a steam world, depending on the exact H$_2$O vs H$_2$ content. The observed trend also implies the presence of a critical temperature (T$_{\rm crit}$) dividing H$_2$O-rich sub-Neptunes with a cold trap and low stratospheric H$_2$O abundance and those without a cold trap and high H$_2$O abundance. We note however that a subset of sub-Neptunes may be gas dwarfs, with much lower water abundances, which can result in such planets not following this trend.

{While the value of T$_{\rm crit}$ is not yet known theoretically, an empirical estimate is possible based on the H$_2$O abundance of TOI-270~d with $T_{\rm eq}$ of 352 K. Multiple estimates have been reported in the literature for the H$_2$O abundance of TOI-270~d \citep{benneke_jwst_2024,holmberg_possible_2024}. Both studies find the H$_2$O abundance to be intermediate between K2-18~b and GJ~9827~d as shown in Fig.~\ref{fig:abundance_trends}, albeit with significant uncertainty. The intermediate value would be consistent with the possibility of H$_2$O being present in gas phase only on the dayside atmosphere with a cold trap on the nightside, consistent with expectations for a dark hycean world \citep{madhusudhan_habitability_2021, holmberg_possible_2024}\footnote{{A dark hycean is where the liquid water ocean is present only on the night side of the planet (assuming the planet is tidally locked)}}. Based on this estimate, T$_{\rm crit}$ would be expected to be  $\gtrsim 350$ K. On the other hand, if the true H$_2$O abundance is closer to that of GJ~9827~d then T$_{\rm crit}$ would be expected to be below 350 K and the atmosphere would be more consistent with a mixed H$_2$-H$_2$O envelope as suggested by \citep{benneke_jwst_2024}. Therefore, a more precise H$_2$O estimate of TOI-270~d could help establish the true value of T$_{\rm crit}$.}

\section*{Interiors and Surface Conditions} 
\label{sec:interiors}

The atmospheric compositions are beginning to provide initial constraints on the internal structures and surface conditions of sub-Neptunes. One of the longest-standing questions in the field is about identifying which sub-Neptunes are gas dwarfs, mini-Neptunes, or water worlds. Atmospheric measurements are essential to break the internal structure degeneracy, and to explore the diversity of possible internal structures.

\subsection*{Degeneracies in Internal Structures}
In the absence of atmospheric measurements, a degenerate set of internal structures can usually explain the planetary mass and radius. For example, as shown in Fig.\ref{fig:population}, the masses and radii for the three sub-Neptune targets K2-18~b, TOI-270~d, and GJ~9827~d all lie on the pure H$_2$O curve. While their bulk properties can be explained by interiors composed entirely of H$_2$O, such a prospect is unlikely based on planetary formation models, which require {a} non-negligible rock component in planetary embryos. Conversely, the bulk parameters may be explained by varied mass fractions of a rocky core, H$_2$O mantle and H$_2$-rich envelope, making it a degenerate problem \citep{Rogers2010a}.

As discussed above, the atmospheric compositions of several sub-Neptunes retrieved from JWST observations have revealed either H$_2$-rich or H$_2$O-rich atmospheres. While an H$_2$O dominated atmosphere, e.g. of GJ~9827~d, makes the planet very likely a hot water world (a `steam world'), the presence of an H$_2$-rich atmosphere, e.g. for K2-18~b or TOI-270~d, still allows for degeneracies in their internal structures. For example, the bulk parameters of K2-18~b could be explained by any of the three sub-Neptune scenarios discussed above \citep{Madhusudhan2020, Madhusudhan_chem_2023}, i.e., a gas dwarf, a mini-Neptune, or a hycean world. However, atmospheric constraints on prominent molecules besides H$_2$, such as CH$_4$, CO, CO$_2$ and NH$_3$, are providing a powerful avenue to further resolve this degeneracy \citep{Yu2021, Hu2021, Tsai2021, Madhusudhan_chem_2023}. Such inferences are made using a combination of atmospheric and internal structure models for the different scenarios as discussed below. 

\subsection*{Atmospheric Signatures of Interiors and Surfaces}

Photochemical models have been used to predict expected atmospheric compositions for the three sub-Neptune scenarios with H$_2$-rich atmospheres discussed above: (a) mini-Neptunes, (b) gas dwarfs, and (c) hycean worlds. The relative abundances of chemical species in sub-Neptune atmospheres depend on various factors, including the atmospheric metallicity, internal temperature, incident radiation, dynamical processes and the presence or lack of a surface \cite{Yu2021, Hu2021, Tsai2021, madhusudhan_carbon-bearing_2023, Cooke2024, Rigby_magma_2024}. Here, we briefly review the broad atmospheric predictions for each scenario, focusing on temperate to warm ($T_{\rm eq}$ $\sim$300-600 K) sub-Neptunes for which robust chemical detections have been made with JWST. 

In temperate mini-Neptune atmospheres, with deep H$_2$-rich atmospheres, H$_2$O, CH$_4$ and NH$_3$ are expected to be the dominant molecules containing O, C, and N, respectively \citep{Yu2021, Hu_photo21,Cooke2024}, as discussed in section {\it Atmospheric Processes}. For warmer planets, CH$_4$ and NH$_3$ may be replaced by CO and N$_2$, respectively, due to vertical mixing from deeper hotter regions. CO$_2$ can also be abundant for high metallicity, and CO is expected to be more abundant than CO$_2$, for solar-like elemental ratios. H$_2$O may not be readily observable in very cool atmospheres due to condensation. The retrieved abundances of the warm exo-Neptune GJ~3470~b, as discussed in the section {\it Chemical Disequilibrium and Atmospheric Extent}, as well as the solar system ice giants are consistent with this composition. Mini-Neptunes are expected to follow the same abundance patterns as ice giants but with higher metallicities, {if the mass-metallicity relation from the solar system were to be extrapolated.}  However, it is possible that several sub-Neptunes could also be gas dwarfs with rock dominated interiors and low volatile abundances in the atmospheres and hence not match this relation.

The expected atmospheric composition for gas dwarfs is similar to the mini-Neptune scenario, with the common aspect being their deep H$_2$-rich atmospheres \citep{Rigby_magma_2024}. For gas dwarfs with high enough temperatures and thick enough atmospheres the rocky surface below could be molten, leading to melt-atmosphere interactions that could affect the observed abundances, while still predicting CO $>$ CO$_2$ \cite{Schlichting2022}. An additional consideration is the possibility of a reduced mantle which can act as a sink for nitrogen \citep{Daviau2021, Suer2023}. However, a molten surface or N-depletion is not necessarily always feasible for temperate sub-Neptunes, and assuming so \citep{shorttle_distinguishing_2024} can result in unphysical solutions \citep{Rigby_magma_2024}. Another key difference is that atmospheric metallicities for gas dwarfs {with just a rocky interior and an H$_2$-rich envelope}, would be expected to be lower than {those of} mini-Neptunes which are {predicted} to have high elemental abundances due to accreted ices during formation \cite{Venturini2020A&A}. 

The expected atmospheric compositions for hycean planets represent a significant departure from both the mini-Neptune and gas dwarf scenarios. Given their thin H$_2$-rich atmospheres their observable compositions are significantly affected by the presence of an ocean surface below which curtails the recycling of key molecules such as CH$_4$ and NH$_3$ from the deeper atmosphere which is in thermochemical equilibrium to the upper, observable, atmosphere where they can be photochemically destroyed \citep{Yu2021, Hu2021, Tsai2021, Madhusudhan_chem_2023}. The prominent molecules in such atmospheres are expected to be CO$_2$, that is photochemical in origin or outgassed from the ocean, and/or CH$_4$, that survives photochemistry or is biogenic. A key indicator of a hycean atmosphere is CO$_2$ $>$ CO and  depletion of NH$_3$ owing to both photochemical destruction and its high solubility in the ocean \cite{Hu2021, Tsai2021_vulcan, Madhusudhan_chem_2023}. While H$_2$O may be depleted in temperate hycean atmospheres where a cold trap may be present \citep{Madhusudhan_chem_2023}, it could be observable in dark hycean atmospheres with warmer daysides \citep{madhusudhan_habitability_2021}. Finally, for an inhabited {h}ycean world, the atmosphere may contain observable quantities of biogenic gases such as DMS, CS$_2$, OCS, CH$_3$Cl, and N$_2$O  \citep{madhusudhan_habitability_2021}.

\begin{figure}[t!]
	\centering \includegraphics[width=1\columnwidth]{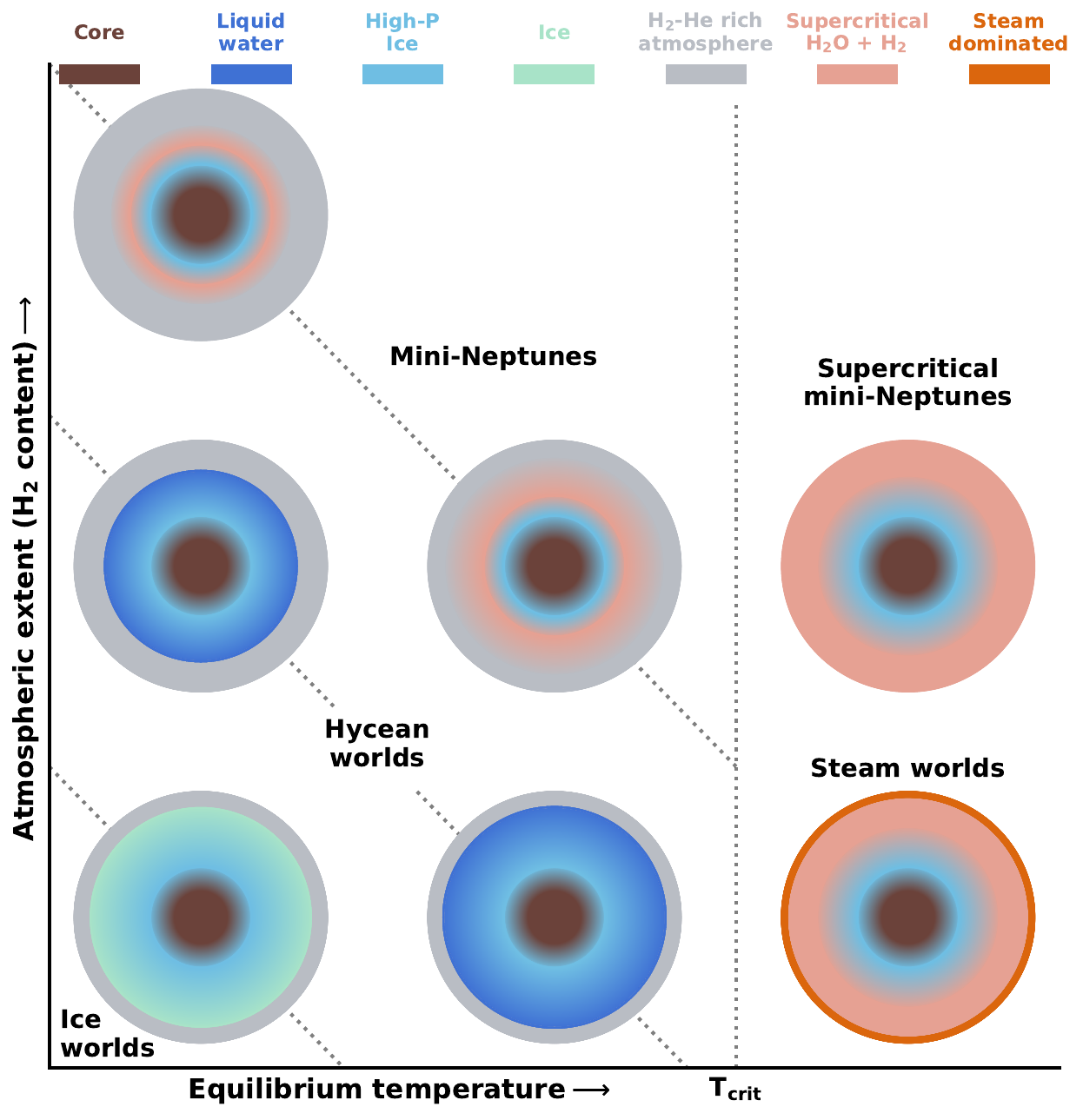}
    \caption{A classification scheme for volatile-rich sub-Neptunes. The schematic compares different types of sub-Neptunes with H$_2$ and H$_2$O dominated atmospheres as a function of the equilibrium temperature and H$_2$ content. \textbf{Ice worlds}: Cold sub-Neptunes where any water is frozen in a layer of ice atop high-pressure ice, with a {thin} H$_2$-dominated atmosphere \citep{Nixon21_ocean}. \textbf{{H}ycean worlds}: Warmer than ice worlds, {h}ycean worlds have a thin H$_{2}$-rich atmosphere overlying an ocean of liquid water, which itself is above {a layer of} high-pressure ices. These planets have the potential to support {habitable conditions} \citep{madhusudhan_habitability_2021}. \textbf{Mini-Neptunes}: {Sub-Neptunes with a deep H$_2$-rich atmosphere followed by a mixed} H$_2$O-H$_2$ {supercritical layer below} before reaching a layer of high-pressure ice and then a core \citep{Nixon21_ocean, Hu_photo21}. \textbf{Steam worlds}: {Water worlds with H$_2$O dominated atmospheres.} At equilibrium temperatures above $\textrm{T}_\textrm{crit}$, a high proportion of H$_2$O exists in the atmosphere, with a supercritical mixed-envelope at higher pressures. \textbf{Supercritical mini-Neptunes}: These planets are hotter than $\textrm{T}_\textrm{crit}$, and contain interiors where H$_2$ gas is highly soluble in supercritical H$_2$O \citep{Nixon21_ocean, benneke_jwst_2024}. This mixed-envelope {would lie over} a layer of high-pressure ice above the core. \textbf{Legend}: The legend at the top {shows the different compositions corresponding to the colors in the schematic, including the core, different phases of H$_2$O and envelope composition; high-P ice refers to high-pressure ice.} Note that the core has further distinct layers but we do not label them here. Sharp phase transitions are shown for ice-atmosphere and ocean-atmosphere transitions. Other phase transitions are marked by a gradient, as they may not be sharp phase transitions \citep{Guillot_planets}.}
    \label{Sub-Neptune schematic figure}
    \vspace{-5mm}
\end{figure}

\subsection*{Classification of Volatile-rich Sub-Neptune Interiors} 
\label{sec:classification}
An emerging picture is that sub-Neptunes with significant volatile inventories (i.e. of H$_2$ and H$_2$O), including mini-Neptunes and {h}ycean worlds, may represent a continuum in internal structures as a function of incident irradiation and atmospheric extent \citep{Nixon21_ocean, rigby_ocean_2024, benneke_jwst_2024}. A schematic of a classification in this two-dimensional space is shown in Fig.~\ref{Sub-Neptune schematic figure}, with $T_{\rm eq}$ representing the irradiation and {atmospheric extent representing the amount of H$_2$ available which in turn affects the total atmospheric pressure. For a H$_2$ dominated atmosphere the total atmospheric pressure (P$_{\rm atm}$) would be close to the partial pressure of H$_2$ (P$_{\rm H_2}$) with minor contributions from other gases.} For a thin atmosphere, e.g. P$_{\rm H2}$ $\sim$ 1 bar, low temperatures {($T_{\rm eq}$ $\lesssim$ 250 K)} can lead to an icy world with a dry H$_2$-rich atmosphere over an icy surface. For intermediate temperatures, $T_{\rm eq}$ $\sim$ 250-T$_{\rm crit}$ K, the atmosphere may still be depleted in H$_2$O due to a cold trap but the surface {can} be warm enough to sustain liquid water, resulting in a {h}ycean world. At still higher temperatures, $T_{\rm eq}$ $>$ T$_{\rm crit}$, a cold trap would no longer be possible, resulting in a steam world where H$_2$O dominates the atmospheric composition, with supercritical H$_2$O in the interior acting as a large sink for H$_2$, given the high solubility of H$_2$ in supercritical H$_2$O \citep{Soubiran2015, Nixon21_ocean, gupta2025miscibility}. {As discussed above, current empirical 
estimates suggest a T$_{\rm crit}$ $\gtrsim$ 350 K.} 

For a given $T_{\rm eq}$, the surface temperature below the atmosphere increases with total pressure. For intermediate pressures, P$_{\rm H_2}$ between approximately one and a few tens of bar, at low $T_{\rm eq}$ the atmosphere {may} still sustain a hycean world {for a high enough albedo}. For higher temperatures, the atmosphere would {continue to} be H$_2$ dominated for $T_{\rm eq}$ $<$ T$_{\rm crit}$ while a cold trap {still persists. A cold trap in such a scenario,} either throughout the atmosphere or only partially such as on the cooler night side, {may} still prevent substantial H$_2$O in the atmosphere. However, the surface and interior would be in supercritical phase followed by high-pressure ice at depth, resulting in a mini-Neptune structure. For, $T_{\rm eq}$ $>$ T$_{\rm crit}$, the lack of a cold trap would result in a high H$_2$O abundance in the atmosphere and a supercritical mixed envelope of H$_2$ and H$_2$O \citep{Nixon21_ocean, benneke_jwst_2024}, which we refer to as a supercritical mini-Neptune. This state is equivalent to the mixed envelope \citep{Nixon21_ocean} or miscible envelope sub-Neptune \citep{benneke_jwst_2024} scenario referred to in previous works. We use the term supercritical to distinguish this state from mixing that can occur in planetary interiors by other means, e.g., through dynamical processes. For higher P$_{\rm H_2}$, a {h}ycean world may no longer be possible, and the internal structure would be that of a mini-Neptune for $T_{\rm eq}$ $<$ T$_{\rm crit}$ or a supercritical mini-Neptune otherwise. For supercritical mini-Neptunes, the amount of H$_2$O in the atmosphere depends on the relative mass fractions of H$_2$ and H$_2$O available. Planets with {lower H$_2$O mass fractions in the interior, possible due to low ice accretion during formation,} would be expected to have lower H$_2$O abundances in the atmosphere.

\subsection*{JWST Constraints on Sub-Neptune Interiors} The retrieved atmospheric abundances provide initial constraints on the internal structures as discussed above. For K2-18~b, the abundances of CH$_4$ and CO$_2$ and non-detections/upper-limits of CO and NH$_3$ {are} consistent with prior predictions for a {h}ycean world \citep{Hu2021, Tsai2021, madhusudhan_carbon-bearing_2023}. The constraint on the photospheric temperature and non-detection of water are also consistent with the presence of a cold trap expected at those temperatures \citep{Madhusudhan_chem_2023}. Recent studies investigating alternate scenarios highlight the challenges in explaining the retrieved abundances without a {h}ycean scenario. Models of the mini-Neptune scenario with a deep H$_2$-rich atmosphere \citep{wogan_jwst_2024, Yang_Hu_24, Huang_24_MN} are unable to explain the CO/CO$_2$, NH$_3$ and/or H$_2$O constraints \citep{Cooke2024}. Other studies have highlighted the importance of self-consistent and physically plausible models. For example, Ref \citep{Luu_supercritical}  considered a supercritical ocean to explain the CH$_4$/CO$_2$ ratio and CO non-detection without considering the effect of photochemistry or nitrogen chemistry. Similarly, another study considered the possibility of magma oceans to explain the NH$_3$ depletion \citep{shorttle_distinguishing_2024} while being inconsistent with the CO/CO$_2$ constraints, planet mass and/or bulk density, and other factors \citep{Rigby_magma_2024}. 

The sub-Neptune TOI-270~d has very similar atmospheric features to K2-18~b (see Fig.~\ref{fig:spectra}), with CO$_2$ and CH$_4$ detected, and also H$_2$O, but not NH$_3$ or CO. This indicates that TOI-270~d could be a dark {h}ycean world, without a cold trap on the day side, but with one on the night side \cite{madhusudhan_habitability_2021, holmberg_possible_2024}. A separate analysis of the data detected the same molecules but with significantly higher abundances, by $\sim$1 dex, indicating a higher metallicity atmosphere \citep{benneke_jwst_2024}. The abundances were used to infer a supercritical mini-Neptune with no liquid water surface. Both interpretations are likely plausible for the corresponding retrieved abundances, considering the dark {h}ycean scenario is also expected to be supercritical on the dayside. However, more accurate observations and establishing the cause of the differences between the two analyses can help resolve the debate.

Finally, the hotter and smaller exoplanet, GJ~9827~d, with a high abundance of H$_2$O  in its atmosphere, is consistent with predictions for a steam world \citep{piaulet2024}. The observed composition indicates an H$_2$O-rich atmosphere over a mixed H$_2$-H$_2$O envelope, due to supercritical H$_2$O, with a rocky mantle/core at depth. The source of the water could be primordial accretion or magma–atmosphere interactions \citep{piaulet2024, kite_atmosphere_2020, Schlichting2022}. The bulk density of the planet precludes a predominantly rocky interior as shown in Fig.~\ref{fig:population}. Sulfur and carbon-bearing species, if present, may inform whether the inferred water stems from the accumulation of icy materials or from volatile-deficient formation with subsequent interior–atmosphere evolution \citep{piaulet2024, Yang_Hu_24}. 

\section*{Habitability and Biosignatures} 

Exoplanetary systems allow a wide range of possible environments that can exist within the habitable zone. Three types of temperate exoplanets have been considered to be conducive for habitable conditions: (a) {r}ocky worlds, (b) {o}cean worlds, and (c) {h}ycean worlds. {While rocky planets have traditionally been the primary focus in the search for habitability and biosignatures on exoplanets, temperate sub-Neptunes that can be ocean worlds or hycean worlds are emerging as promising avenues in recent years.}

Habitable rocky worlds, which include temperate Earth-like planets and super-Earths{,} have densities consistent with interiors that are entirely rocky. Atmospheric observations of habitable rocky planets orbiting Sun-like (G dwarf) stars, which is the primary motivation for the Habitable Worlds Observatory (HWO) in the future \citep{Nat_Acad_HWO}, are beyond the capability of JWST. However, such planets around small nearby M dwarfs are accessible to JWST with significant investment of observing time. Recent JWST observations of several terrestrial-size planets in the TRAPPIST-1 system \citep{gillon2017} have highlighted the significant challenges resulting from spectral contamination from the very active late M dwarf star as well as the planets potentially lacking significant atmospheres \citep{greene2023,Lim2023,zieba2023}. An additional challenge is the small number of known transiting planets around nearby bright M dwarfs, though this could change with new planet detections, such as the recently discovered Gliese 12~b \citep{Dholakia2024, Kuzuhara_12b}.

Ocean worlds are temperate water-rich planets with ocean-covered surfaces and terrestrial-like atmospheres \citep{leger_new_2004}. Such planets can have a wide range of water mass fractions, including terrestrial-like rocky interiors but with more water than on Earth, e.g. 10-1000$\times$ Earth oceans \citep{Wordsworth2013, Kite2018}, as well as planets with over 50\% of their mass in water. While many known sub-Neptunes may be water worlds \citep{Zeng2019, Luque2022} most are likely steam worlds that are too hot to be habitable. Recent JWST observations have identified a potential temperate ocean world, LHS~1140~b, inferred by its spectrum being inconsistent with a H$_2$-rich atmosphere, indications of a high MMW atmosphere, and the bulk density being too low for an Earth-like interior \citep{Cadieux2024b}{.}

Hycean worlds, as discussed above, are temperate water-rich planets with ocean covered surfaces and H$_2$-rich atmospheres \citep{madhusudhan_habitability_2021}. Their lower densities, and hence larger radii, and larger atmospheric scale heights make them more conducive to detection and atmospheric characterization compared to rocky or ocean worlds of similar mass. The greenhouse warming due to the H$_2$-rich atmosphere significantly expands the classical terrestrial HZ thereby increasing the available targets for atmospheric observations \citep{madhusudhan_habitability_2021}; over a dozen promising {h}ycean candidates are known around nearby M dwarfs. As discussed in the above section, JWST observations and the retrieved atmospheric properties of the habitable zone sub-Neptune K2-18~b are consistent with prior predictions for a {h}ycean world \citep{madhusudhan_carbon-bearing_2023}. Subsequent JWST observations of a second sub-Neptune TOI-270~d also indicate the possibility of a dark {h}ycean world, albeit the retrieved properties are presently debated \citep{holmberg_possible_2024, benneke_jwst_2024}. 

Overall, {h}ycean worlds present a promising avenue in the search for habitable conditions with JWST, both in terms of observability and the number of {potential} targets available. However, open questions remain about the interpretation of the atmospheric observations, as discussed above, and the potential of current {h}ycean candidates for sustaining habitable conditions. When considering a planet's climate, the Bond albedo and atmospheric {extent} determine whether a temperate sub-Neptune can be a {h}ycean world. For K2-18b, an adequate albedo (e.g.$\sim$0.5-0.6) is required to maintain a liquid water surface \citep{piette_temperature_2020, leconte_3d_2024}; zero-albedo models predict temperatures too hot to be habitable \citep{innes_runaway_2023, Pierrehumbert2023, piette_temperature_2020}. While the observed transmission spectrum provides a 3$\sigma$ evidence for clouds/hazes at the terminator \citep{madhusudhan_carbon-bearing_2023}, with the retrieved haze properties consistent with the required theoretical values \citep{leconte_3d_2024}, the dayside albedo is still unconstrained. However, considering the range of Bond albedos in the solar system planets with significant atmospheres is between $\sim0.3-0.8$, including {0.29 for Earth \citep{Stephens_EarthAlbedo2015}, 0.50 for Jupiter \citep{li2018less} and 0.76 for Venus \citep{Haus_2016_albedo}}, the required Bond albedo for K2-18~b may not be unusual \citep{Cooke2024}. It is also presently unknown if the required thin/moderate atmosphere, with total pressure below{ $\sim 10$ bar} depending on the albedo, can be retained on K2-18~b given the susceptibility to atmospheric loss. Finally, while the current chemical abundances are most consistent with a {h}ycean world scenario, it remains to be seen if new atmospheric processes not considered previously might explain the observed abundances without the need for a {h}ycean scenario. More precise observations and theoretical studies in the future could help answer these questions. 

{
Recent studies have explored the nature and detectability of possibile biosignatures on different types of habitable exoplanets with JWST. For terrestrial-like atmospheres, the detection of ozone (O$_3$) on a TRAPPIST-1 planet would require several tens of transits, over hundreds of hours, with JWST \citep{Barstow2016}. Habitable sub-Neptunes, such as hycean worlds, are more favorable for biosignature detectability due to their larger sizes and lighter atmospheres than rocky planets \cite{madhusudhan_habitability_2021}. Some of the prominent terrestrial biosignatures such as O$_2$ or O$_3$ are unlikely to be abundant in H$_2$-rich atmospheres and others such as CH$_4$ may have abiotic false positives. However, several of the secondary and less abundant biosignatures on Earth may be good biosignatures on hycean worlds. These include molecules such as dimethyl sulfide (DMS), CH$_3$Cl, OCS, and CS$_2$, some of which were also known to be present in the H$_2$-rich atmosphere of the early Earth \citep{Pilcher2003,DomagalGoldman2011}. Theoretical studies have demonstrated that such biosignature molecules are both feasible in significant abundances and detectable with modest amounts of JWST time for super-Earths with H$_2$-rich atmospheres \citep{Seager2013b} and hycean worlds orbiting nearby M dwarfs \citep{madhusudhan_habitability_2021, Tsai24_Bio}. A very tentative inference of DMS was reported using recent JWST observations of K2-18 b \citep{madhusudhan_carbon-bearing_2023}. While it remains to be seen if future observations confirm the finding, these initial observations demonstrate the potential of JWST to detect such molecules in sub-Neptune atmospheres.}

\section*{Summary and Future outlook}

For the past decade prior to JWST, the sub-Neptune regime had been a formidable frontier of exoplanet science. Within its first two years, JWST has already unveiled stunning glimpses of this exotic landscape with unprecedented detail. Robust chemical detections have been reported for several sub-Neptunes, starting with the habitable zone planet K2-18~b and followed by warmer planets TOI-270~d and GJ 9827~d. These results are already providing insights into atmospheric processes and interiors/surface conditions, as well as population-level trends motivating a uniform classification scheme for volatile-rich sub-Neptune interiors. Finally, these insights have opened a new avenue in the search for life, through ocean worlds and {h}ycean worlds. 

These observations have {taught} important lessons both on sub-Neptune properties as well as their observability. First, the chemical detections across the sub-Neptunes observed with JWST suggest a high likelihood for water-rich interiors \citep{madhusudhan_carbon-bearing_2023, Yang_Hu_24, benneke_jwst_2024, piaulet2024}. They also provide empirical evidence that temperate planets with H$_2$-rich atmospheres are more readily observable than hotter planets with equivalent bulk properties due to the possibility of high H$_2$O abundances and, hence, high MMW in the latter, {e.g. for planets such as GJ 9827~d}. {Second, temperate sub-Neptunes below $\sim$400 K, such as K2-18~b and TOI-270~d, also seem to have less spectral attenuation due to clouds/hazes compared to their warmer counterparts. Third, atmospheric spectroscopy of sub-Neptunes orbiting active M dwarfs can be severely impacted by stellar heterogeneities and flares \cite{May2023, Cadieux2024b}. Planets orbiting mid-M dwarfs with low-moderate activity are more conducive for such observations. Finally, prominent molecules such} as CH$_4$, CO$_2$ and H$_2$O in temperate sub-Neptune atmospheres are within reach of JWST, often requiring minimal observing time for optimal targets. {Furthermore}, potential biomarker molecules are {also} detectable in {promising sub-Neptune} atmospheres with JWST, including species like DMS, CS$_2$, CH$_3${Cl}, N$_2$O and OCS \citep{Seager2013b, madhusudhan_habitability_2021, Tsai24_Bio}. 

All the sub-Neptunes with significant atmospheric detections to date show signs of chemical disequilibrium. The inferred abundances also provide initial evidence for diverse internal structures, from steamy water worlds like GJ 9827~d \citep{piaulet2024} to possible hycean worlds like K2-18~b \citep{madhusudhan_carbon-bearing_2023}. It is important to note that these are still very early inferences and further observations and theoretical efforts are needed for more robust characterisation of such planets. Nevertheless, these early insights are already motivating the atmospheric classification of volatile-rich sub-Neptunes as a function of their irradiation and bulk properties, spanning a continuum between hycean worlds, steamy worlds, mini-Neptunes and supercritical mini-Neptunes. They also pave the way for constructing the mass-metallicity relation in the sub-Neptune regime with implications for planetary formation and evolution in this uncharted territory. Ultimately, however, the high sensitivity of JWST observations necessitate robust data reduction and analysis procedures, high-fidelity atmospheric retrieval frameworks and accurate self-consistent interior/atmospheric models for a comprehensive understanding of the sub-Neptune population.

The next steps {on the sub-Neptune frontier} with JWST are in two directions: wider observations {of a broader range of targets} enabling population-level studies, and deeper characterization of promising sub-Neptunes.  Both approaches are essential to answer major open questions in this frontier: How do the atmospheric compositions of sub-Neptunes vary across the bulk properties between terrestrial planets and ice giants? How diverse are the internal structures of sub-Neptunes and what determines the transition between super-Earths and {sub}-Neptunes? Finally, what are the prospects for habitabilty and biosignatures in the sub-Neptune regime? {The} present work highligh{ts} the {major} advances that are happening in our understanding of sub-Neptunes with JWST. These insights, however, are but a glimpse of what vistas await in the JWST era. Further observations promise to solve longstanding mysteries about this exotic {yet} ubiquitous planetary class. This quest may even bring us face to face with one of humanity's earliest and most primal questions: are we alone in the universe? 

\showmatmethods{} 

\acknow{N.M. and M.H. acknowledge support from STFC Center for Doctoral Training (CDT) in Data Intensive Science at the University of Cambridge (grant No. ST/P006787/1). N.M. and S.C. acknowledge support from the UK Research and Innovation (UKRI) Frontier Grant (EP/X025179/1, PI: N. Madhusudhan). N.M. and G.J.C. acknowledge support from the Leverhulme Centre for Life in the Universe. This research has made use of the NASA Exoplanet Archive, which is operated by the California Institute of Technology, under contract with the National Aeronautics and Space Administration under the Exoplanet Exploration Program. {We thank the editor and reviewers for their valuable comments.}}

\showacknow{} 

 \newcommand{\noop}[1]{}

\end{document}